\documentclass[
reprint,
superscriptaddress,
frontmatterverbose, 
showpacs,
preprintnumbers,
nofootinbib,
amsmath,
amssymb,
prd,
floatfix,
twocolumn
]{revtex4-2}

\usepackage{soul}
\usepackage[inline]{enumitem}
\usepackage[utf8]{inputenc}
\usepackage[normalem]{ulem}
\usepackage{graphicx}
\usepackage{dcolumn}
\usepackage{bm}
\usepackage{color}
\usepackage[dvipsnames]{xcolor}
\usepackage[
    colorlinks = true,
    linkcolor = BlueViolet,
    anchorcolor = purple,
    citecolor = purple,
    filecolor = purple,
    urlcolor = BlueViolet]{hyperref}
    
\usepackage{url}
\usepackage{xspace}
\usepackage{slashed}
\usepackage{multirow,bigstrut}
\usepackage{mathrsfs} 
\usepackage{relsize,amsmath}
\usepackage[geometry]{ifsym}
\usepackage{amssymb}
\usepackage{pifont}
\usepackage{physics}
\usepackage[nameinlink,capitalize]{cleveref} 
\usepackage{comment}

\usepackage{rotating}
\usepackage{ragged2e}
\usepackage{lipsum}

\usepackage{fontawesome} 
\definecolor{blue-violet}{rgb}{0.33, 0.17, 0.89}

\usepackage{siunitx}

\begin{document}

\preprint{\hfill UMN-TH-4529/26}
\title{A New Source of Millicharged Particles:\\ Secondary Showers in the LHC Forward Absorber}

\author{Jyotismita Adhikary}
\email{jyotismita.adhikary@ncbj.gov.pl}
\affiliation{National Centre for Nuclear Research, Pasteura 7, Warsaw, 02-093, Poland}

\author{Peiran Li} 
\email{li001800@umn.edu}
\affiliation{School of Physics and Astronomy, University of Minnesota, Minneapolis, MN 55455, USA}

\author{Zhen Liu}
\email{zliuphys@umn.edu}
\affiliation{School of Physics and Astronomy, University of Minnesota, Minneapolis, MN 55455, USA}

\author{Sebastian Trojanowski}
\email{sebastian.trojanowski@ncbj.gov.pl}
\affiliation{National Centre for Nuclear Research, Pasteura 7, Warsaw, 02-093, Poland}
\affiliation{Astrocent, Nicolaus Copernicus Astronomical Center Polish Academy of Sciences, ul. Rektorska 4, 00-614, Warsaw, Poland}

\author{Azam Zabihi}
\email{azabihi@camk.edu.pl}
\affiliation{Astrocent, Nicolaus Copernicus Astronomical Center Polish Academy of Sciences, ul. Rektorska 4, 00-614, Warsaw, Poland}


\begin{abstract}
Millicharged particles (mCPs) are a well-motivated target for far-forward searches at the Large Hadron Collider. We identify and quantify a significant new source of these particles: secondary production in hadronic and electromagnetic showers initiated by energetic neutral particles striking the TAXN absorber. By combining Monte Carlo simulations with \texttt{Geant4}-based modeling
, we show that these secondary cascades yield a substantial mCP flux that complements the primary production from the interaction point. For the proposed FORMOSA detector, this contribution can enhance the expected signal yield by approximately $50\%$ for $m_\chi \lesssim 0.1~\textrm{GeV}$. Our results demonstrate that secondary production in downstream infrastructure is an essential ingredient for realistic sensitivity projections and new-physics searches at the High-Luminosity LHC. The simulated secondary spectra are made publicly available to facilitate future forward physics studies.\footnote{\label{ft:repository}To facilitate future research into these novel experimental signatures, we provide the simulated secondary hadron and electromagnetic spectra generated for this work in a public repository available at \url{https://github.com/JyotismitaAdhikary/Secondary_Showers-TAXN}.}
\end{abstract}

\maketitle

\section{Introduction}

The Large Hadron Collider (LHC) has delivered major insights into fundamental interactions, including the discovery of the Higgs boson~\cite{CMS:2012qbp,ATLAS:2012yve}. Looking ahead, future LHC data are expected to further extend this scientific program, motivating both detector upgrades and new experimental strategies. In parallel, many novel signatures and search concepts have been proposed for physics beyond the Standard Model (BSM) and for precision Standard Model (SM) measurements, with applications extending into the High-Luminosity LHC (HL-LHC) era; see, for example, Refs.~\cite{Alimena:2019zri,Alimena:2021mdu,CMS:2024zqs}.

A notable recent example of such a new strategy is a far-forward research program at the LHC, which began during the ongoing Run 3 data-taking period. This corresponds to the FASER experiment~\cite{FASER:2018ceo,FASER:2018eoc} dedicated to searching for light and long-lived particles (LLPs)~\cite{Feng:2017uoz} and to neutrino detectors FASER$\nu$~\cite{FASER:2019dxq} and SND@LHC~\cite{SHiP:2020sos}, which have recently directly observed collider neutrinos for the first time~\cite{FASER:2023zcr,SNDLHC:2023pun}. Beyond Run 3, far-forward studies will be continued in FASER and SND@LHC in the LHC Run 4 data-taking period~\cite{Boyd:2882503,FASER:2025myb,SNDLHC:2026why}, and can further extend to the HL-LHC era with the proposed Forward Physics Facility (FPF)~\cite{Anchordoqui:2021ghd,Feng:2022inv,FPF:2025bor}. 

\begin{figure*}
\centering
\includegraphics[width=1\textwidth]{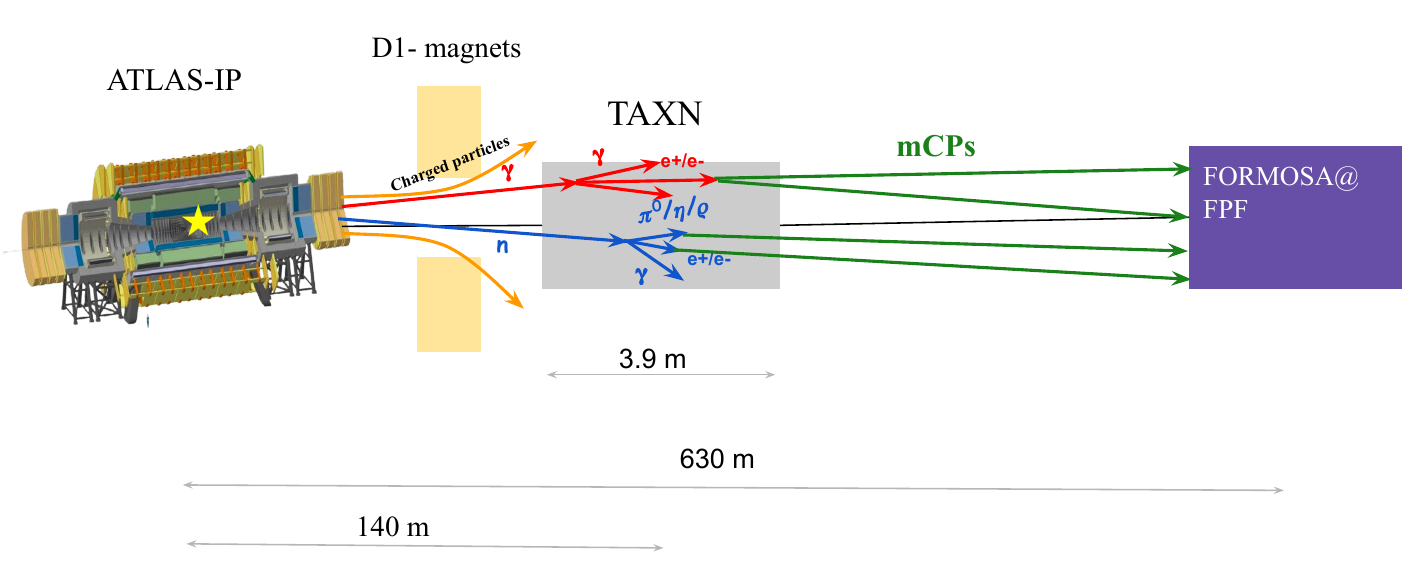}
\caption{Schematic illustration of secondary mCP production at TAXN. Primary photons originating at the IP initiate electromagnetic showers, whereas primary neutrons induce both electromagnetic and hadronic cascades. Secondary particles emerging from both shower types subsequently produce mCPs at TAXN.
}
\label{fig:Schematic}
\end{figure*}

In particular, the search for ionization signals from beyond the Standard Model (BSM) millicharged particles (mCPs) in the forward kinematic region of the LHC has been proposed~\cite{Foroughi-Abari:2020qar}, and the dedicated FORMOSA detector is currently under design~\cite{Citron:2025kcy}. These postulated mCPs, carrying a tiny electric charge $Q = \epsilon e$ with $\epsilon\ll 1$, are predicted in various extensions of the SM. They can arise in the hidden sector due to kinetic mixing between a massless dark photon and the SM hypercharge gauge boson~\cite{Galison:1983pa,Holdom:1985ag}. Their existence may also be linked to charge quantization in the framework of Grand Unified Theories (GUTs)~\cite{Okun:1983vw,Brahm:1989jh}. Since mCPs are predicted to be stable, they could naturally constitute a fraction of dark matter (DM) in the universe, requiring novel experimental probes~\cite{Essig:2011nj,Kouvaris:2014lpa,Essig:2015cda,Essig:2017kqs,Bhoonah:2018wmw,Budker:2021quh,Berlin:2023zpn,Berlin:2025btf,Berlin:2025hjs}. While cosmological bounds can constrain the mCP fraction of DM~\cite{dePutter:2018xte,Kovetz:2018zan,Berlin:2022hmt}, such bounds depend on the assumption about early universe evolution~\cite{Adshead:2022ovo,Gan:2023jbs}. In contrast, searches for laboratory-produced mCPs provide constraints that are independent of the cosmological history~\cite{Haas:2014dda,Magill:2018tbb,Kelly:2018brz,Choi:2020mbk,Kim:2021eix,Kling:2022ykt,Kalliokoski:2023cgw,MammenAbraham:2024gun,Essig:2024dpa,Eberl:2025kfm}.

The FORMOSA experiment is proposed to search for mCP ionization signals, following the similar concept of the milliQan detector placed in the central region of the LHC~\cite{Ball:2016zrp}. The detector could be placed either within the FPF tunnel or in an alternative location in one of the existing caverns near the LHC beam, which also hosts the current FORMOSA demonstrator~\cite{Citron:2025kcy}. FORMOSA will utilize an expected large flux of forward-going mCPs produced in hadron decays or direct production in $pp$ collisions in the ATLAS interaction point (IP) at the LHC. 

Although the LHC is primarily a collider, it can also operate at specific fixed-target modes. A notable example is the SMOG system installed in the LHCb experiment to study hadron production on nuclear targets at rest~\cite{LHCb:2018jry,LHCb:2018ygc}, while similar searches have also been proposed for the future~\cite{Barschel:2020drr}. Interestingly, however, the LHC effectively functions as a beam-dump experiment even without any modifications to the existing infrastructure. In particular, the large and collimated fluxes of high-energy neutral particles produced in $pp$ collisions that travel down the beam pipe eventually hit neutral particle absorbers without being deflected by the LHC magnets. This idea has been utilized to propose a FASER search for axion-like particles coupled to photons, which could be produced via the Primakoff process in high-energy photon scatterings, $\gamma N \to a N$, in the target absorber for neutrals (TAN), located about $140~\textrm{m}$ from the IP~\cite{Feng:2018pew}; see~\cite{FASER:2024bbl} for recent search results.

In addition to the initial interaction with the absorber, new particles can appear in subsequent interactions within the target, both in electromagnetic (EM) and hadronic showers~\cite{Dev:2023zts}. These secondaries correspond to lower interaction energies, but their production rates remain substantial for low-mass species. For example, the average energy of a photon incident on the TAN within the angular acceptance of the FASER detector is of order $E_\gamma\sim 800~\textrm{GeV}$. When hitting the absorber, each such photon can generate hundreds of secondary photons, electrons, and positrons with $E_{\gamma,e^\pm}>\textrm{GeV}$. Similarly, for $\sim \textrm{TeV}$-scale incident forward neutrons at the LHC, the fraction of the neutron energy transferred to the EM shower is about $70\%$, cf. Groom and Wigman shower parameterizations~\cite{Wigmans:1987cz,Donaldson:1990uu}, producing a similar large enhancement. Notably, this fraction is largely due to decays of secondary neutral pion and eta mesons, which provide numerous additional opportunities to produce BSM particles in the absorber. We provide the corresponding simulated secondary spectra online\footref{ft:repository}.

The main idea of this work is summarized in \cref{fig:Schematic}. In this study, we illustrate how such secondary production in neutron- and photon-induced showers in the TAN absorber is expected to impact the forward mCP flux at the LHC. The millicharged particles are a perfect playground for testing the importance of secondary production effects, since even soft such particles can easily reach far-forward detectors and leave distinct ionization signals. As discussed below, this can result in an even $50-60\%$ enhancement of the predicted mCP interaction rate in FORMOSA, which has not been studied previously. 

This paper is organized as follows. In \cref{sec:secondaries}, we describe the forward particle spectra at the LHC and the modeling of secondary shower development within the TAXN absorber. \cref{sec:mcp_at_LHC} details the various mCP production channels, including both primary interaction point and secondary shower contributions, and outlines the ionization-based detection strategy. Our main results, including the relative importance of secondary production and the projected sensitivity reach, are presented in \cref{sec:results}. Finally, we summarize our findings and provide a future outlook in \cref{sec:conclusion}.

\section{LHC as a beam-dump and forward spectra}
\label{sec:secondaries}
\begin{figure*}
\centering
\includegraphics[width=\columnwidth]{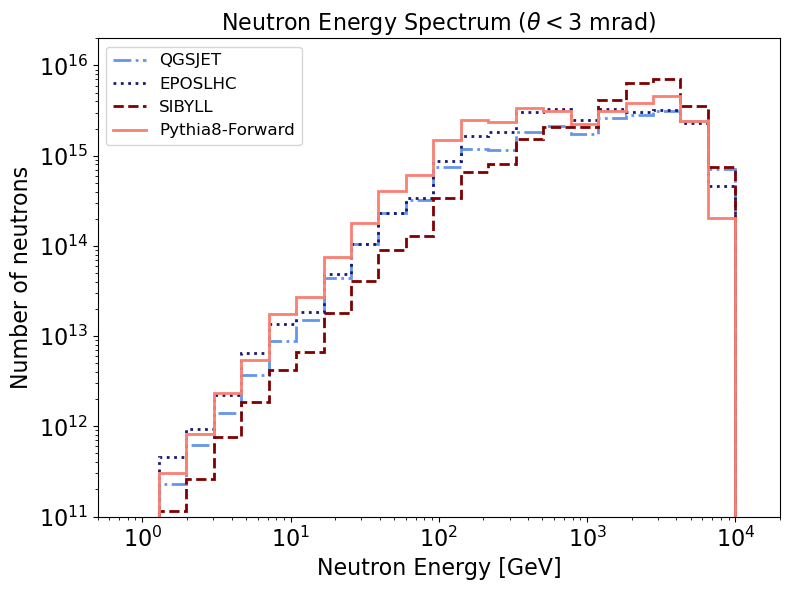}
\hfill
\includegraphics[width=\columnwidth]{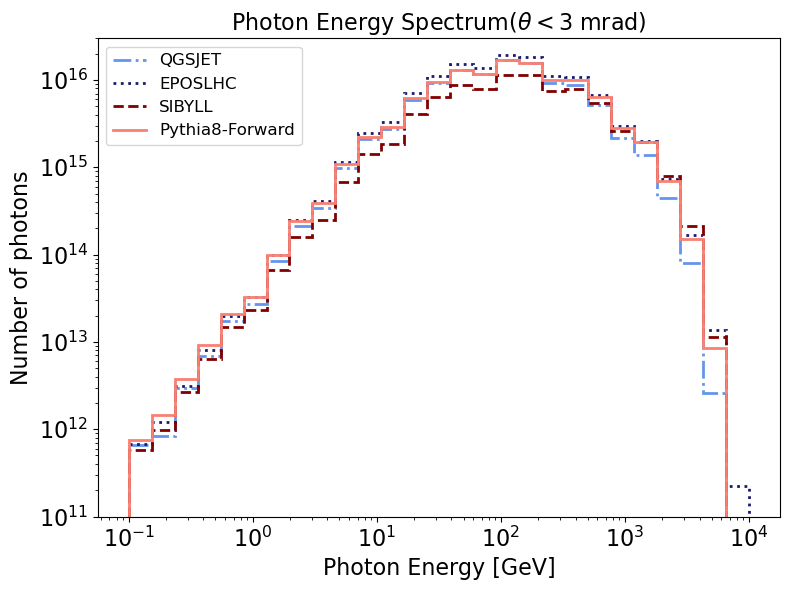}
\caption{Predicted energy spectra of primary forward neutrons (left) and photons (right) produced within $\theta < 3~\text{mrad}$ of the LHC beam collision axis. The results are obtained using various dedicated Monte Carlo event generators, assuming an integrated luminosity of $3~\text{ab}^{-1}$. Different colors represent different generators, while a consistent line style is maintained for all predictions. 
}
\label{fig:primary_spectra}
\end{figure*}
A typical proton collision at the LHC produces tens to hundreds of secondary particles following a transverse momentum distribution that peaks around their mass, $\langle p_T\rangle \sim m$, where $m\lesssim \textrm{several GeV}$ for most secondaries. Such high-energy species, with a total momentum of $p_{\textrm{tot}}\sim \textrm{TeV}$, are strongly collimated along the beam collision axis, exhibiting a tiny characteristic angular spread relative to the beam of $\theta \sim \langle p_T\rangle/p_{\textrm{tot}}\ll 1$. While charged secondaries are effectively deflected by strong LHC magnets, neutral particles travel down the beam pipe until they strike forward particle absorbers that shield the downstream LHC infrastructure.

These particles are primarily photons ($\gamma$) and neutrons ($n$). The former are abundantly produced through the decays of forward-going neutral pions ($\pi^0$) and eta mesons ($\eta$) close to the IP, and measuring their forward spectra provides a window into studying high-energy $pp$ collisions beyond the perturbative regime~\cite{LHCForwardPhysicsWorkingGroup:2016ote}. The production of forward neutrons, $\pi^0$, and $\eta$ are constrained by LHCf data~\cite{LHCf:2014gqm,LHCf:2015nel,LHCf:2015rcj,LHCf:2017fnw,LHCf:2018gbv,LHCf:2020hjf,Piparo:2023yam}, while other hadrons are indirectly studied by measuring forward neutrinos~\cite{Bai:2020ukz,Kling:2021gos,Bai:2021ira,Bai:2022xad,Maciula:2022lzk,Bhattacharya:2023zei,Buonocore:2023kna,Kling:2023tgr,FASER:2024ykc,FASER:2024hoe}.

These data will help to validate dedicated Monte Carlo (MC) event generators, such as EPOS-LHC~\cite{Pierog:2013ria}, QGSJET~\cite{Ostapchenko:2010vb}, SIBYLL~\cite{Ahn:2009wx,Ahn:2011wt,Riehn:2015oba,Fedynitch:2018cbl}, and a recent forward Pythia tune~\cite{Fieg:2023kld}. In \cref{fig:primary_spectra}, we show a comparison of the energy spectra obtained with these generators, representing forward neutrons and photons produced within an angle of $\theta<3~\textrm{mrad}$ from the beam collision axis.\footnote{While the current FASER detector has a smaller angular acceptance of $\theta < 0.21~\textrm{mrad}$~\cite{FASER:2022hcn}, resulting in a harder spectrum, we adopt a 3 mrad cut to accommodate proposed FASER extensions~\cite{FPF:2025bor}. This choice ensures the inclusion of larger transverse detector geometries and provides sufficient margin to fully capture the impact of secondary production in both EM and hadronic showers.} The plot assumes $3~\textrm{ab}^{-1}$ of integrated luminosity, relevant for the future High-Luminosity era of the LHC (HL-LHC), and was obtained with the \texttt{FORESEE} numerical package~\cite{Kling:2021fwx}.

As can be seen, in each case, the expected forward neutron spectra peak at about the TeV energy scale, while forward photons are softer and peak around a few hundred GeV. This is expected based on the characteristic transverse momentum discussed above, which is larger for neutrons than for the $\pi^0$ and $\eta$ mesons that source forward photons. As a result, a larger total energy is typically required to focus neutrons within a narrow angular spread of $\theta<3~\textrm{mrad}$.

Some differences between the predictions can also be noted. In particular, the neutron spectrum obtained with SIBYLL is marginally harder than in the other cases. Conversely, the Pythia tune predicts a forward neutron spectrum that is slightly more pronounced towards the low-energy tail. In both cases, the predicted neutron flux is relatively enhanced compared to EPOS-LHC. In contrast, no such enhancement is seen for photons, which is related to the parent forward pion and eta meson spectra. As we will discuss below, this translates into a notable increase in the predicted significance of forward mCP production in neutron-induced showers compared to primary production in $\pi^0$ and $\eta$ decays when SIBYLL and the Pythia tune are used instead of EPOS-LHC.

\begin{figure*}
\centering
\includegraphics[width=\columnwidth]{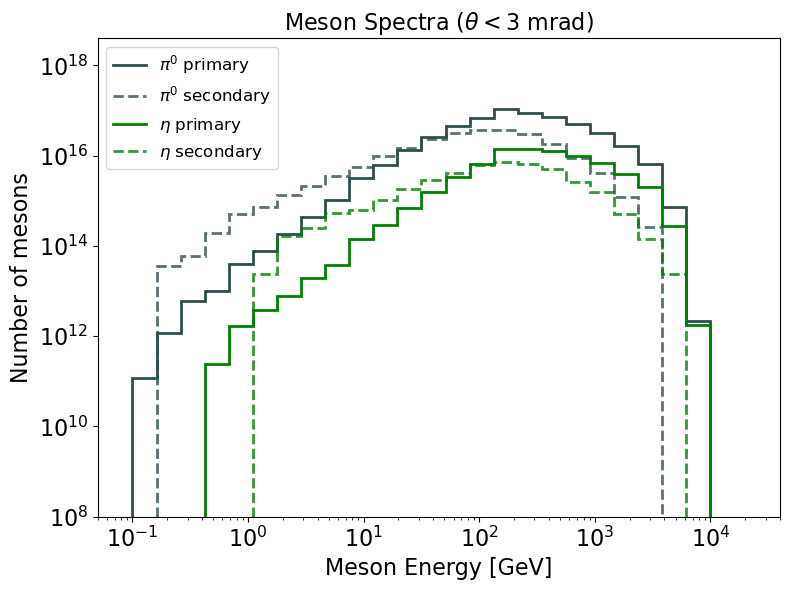}
\hfill
\includegraphics[width=\columnwidth]{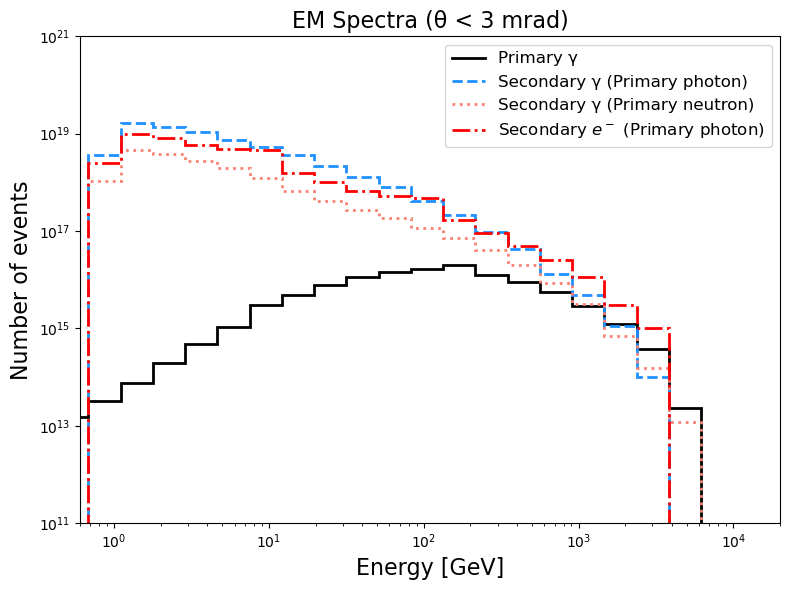}
\caption{Predicted energy spectra of secondary neutral mesons (left) and electromagnetic shower products (right) produced within $\theta < 3~\text{mrad}$ with respect to the LHC beam collision axis. The results are obtained using a forward-tuned Pythia 8 generator, assuming an HL-LHC integrated luminosity of $3~\textrm{ab}^{-1}$. 
}
\label{fig:secondary_spectra}
\end{figure*}

After propagating downstream from the IP, these neutral particles reach the neutral-particle absorber and initiate EM and hadronic showers. In the HL-LHC era, the TAN will be replaced by the new TAXN absorber, located approximately 130 m from the primary IP~\cite{ZurbanoFernandez:2020cco}. For our study, the TAXN is modeled as a solid copper block with dimensions $1~\mathrm{m}\times 1~\mathrm{m}\times 4~\mathrm{m}$.\footnote{At approximately 19.05 m from the IP, forward-going particles first encounter the TA(X)S collimators. In the HL-LHC, the TAXS is planned to have a 60 mm diameter aperture, resulting in an angular coverage of $\theta<1.57~\textrm{mrad}$~\cite{ZurbanoFernandez:2020cco}. Interactions in this absorber effectively shift the secondary production point closer to the IP for larger angles, as showering happens in the TAXS rather than TAXN. We neglect this effect in our analysis below. This simplification is justified as the majority of secondary production relevant to a detector size considered below occurs within the $\theta\lesssim 1.5~\textrm{mrad}$ acceptance, which passes through the TAXS aperture.} We simulate shower development within the TAXN using the \texttt{Geant4} toolkit (version 11.3), employing the \texttt{FTFP\_BERT} physics list~\cite{GEANT4:2002zbu,Allison:2006ve,Allison:2016lfl}.

The resulting EM and hadronic cascades produce a large multiplicity of secondary particles, including photons, $e^\pm$, and neutral mesons. \Cref{fig:secondary_spectra} shows the energy spectra of secondaries produced within the 3 mrad cut, derived from the primary spectra generated with the Pythia 8 forward tune. As shown in the left panel, while secondary $\pi^0$ and $\eta$ mesons are generally softer than their primary counterparts, the hadronic shower contribution becomes dominant at energies $\lesssim 100~\textrm{GeV}$, potentially leading to additional mCP production via rare meson decays. We also note that the production of heavier mesons within the shower is highly suppressed relative to direct production in $pp$ collisions at the IP. Therefore, we focus only on light neutral hadronic secondaries, specifically $\pi^0$, $\eta$, and $\rho^0$ mesons. Given their short lifetimes, these mesons are assumed to decay promptly into final states containing mCPs without requiring further propagation within the TAXN.

The results for EM showers are shown in the right panel of \cref{fig:secondary_spectra}. 
For energies $\lesssim 1~\textrm{TeV}$, the secondary photon fluxes exceed the primary flux, with an enhancement of several orders of magnitude at lower energies. 
Both photon- and neutron-induced showers contribute to secondary photon production, with the photon-induced component generally dominating except at energies approaching a few TeV. 
We also show the spectrum of secondary electrons produced in the EM cascade. The softer spectrum of secondary photons is a direct consequence of the bremsstrahlung process, which preferentially produces low-energy photons, resulting in a steeply falling energy distribution. In contrast, the secondary electron spectrum is harder, reflecting the kinematics of pair production, in which the two leptons carry away nearly the entire energy of the parent photon and thus tend to inherit a significant fraction of it.
Notably, unlike high-energy photons, which are promptly converted into $e^+e^-$ pairs after a relatively short mean free path, electrons and positrons can propagate through the material while losing energy gradually through repeated electromagnetic interactions. 
Their longer trajectories enhance the mCP production inside the TAXN. 
Consequently, both the sizable secondary electron multiplicity and their extended propagation in the absorber must be taken into account when evaluating the EM-shower contribution to the total mCP signal.

All secondary spectra employed in this work, generated using \texttt{Geant4} based on the four aforementioned MC generators, are publicly available online (see footnote 1). While we focus on mCP production here, these datasets may be employed for future studies involving other secondary BSM particle production.

\section{Millicharged particles at the LHC}
\label{sec:mcp_at_LHC}
Millicharged particles arise naturally in dark-sector models containing an additional $U(1)_D$ gauge symmetry with kinetic mixing between the corresponding dark photon and the Standard Model photon~\cite{Holdom:1985ag,Galison:1983pa}. A particularly well-motivated possibility is that dark matter is charged under this dark gauge symmetry and interacts with the visible sector through a dark photon. At low energies, the relevant effective Lagrangian is
\begin{align}
\mathcal{L} \supset -\frac14 F'_{\mu\nu}F'^{\mu\nu}
-\frac{\kappa}{2}F'_{\mu\nu}F^{\mu\nu}
-g_D A'_\mu\bar{\chi}\gamma^\mu \chi ~,
\end{align}
where $A_\mu$ denotes the ordinary photon, $A'_\mu$ the ultralight dark photon, and $\chi$ a dark-sector fermion. After diagonalizing the kinetic term, the dark-sector fermion acquires an effective coupling to the ordinary photon, which can be written as
\begin{align}
    \epsilon e = \kappa g_D ~,
\end{align}
so that $\chi$ behaves as a particle with an effective millicharge $\epsilon e$.

In this work, we study the production of such mCPs in the far-forward region of the LHC, focusing on the proposed FORMOSA detector. We begin with the primary mCP flux from the ATLAS interaction point, which serves as the baseline for the analysis below. We then turn to the secondary contribution from particle showers in the downstream TAXN absorber.

\subsection{mCPs from IP}
\label{sec:mcp_from_ip}

Before discussing the secondary production in the TAXN, we first establish the baseline mCP signal originating directly from the ATLAS interaction point. The forward mCP flux from primary $pp$ collisions, and its relevance for dedicated far-forward detectors, has been discussed in the literature~\cite{Foroughi-Abari:2020qar,Kling:2022ykt}. Here we briefly summarize the production channels at the IP and the simulation procedure adopted in this work. They provide the baseline for comparison with the downstream secondary contribution.

In the mass range of interest, the dominant primary source arises from neutral mesons produced promptly in the far-forward region of $pp$ collisions. In particular, pseudoscalar mesons such as $\pi^0$ and $\eta$ can produce mCPs through the three-body decay
\begin{align}
    M \to \gamma \chi \bar{\chi},
\end{align}
while vector mesons such as $\rho^0$ can contribute through the two-body decay
\begin{align}
    M \to \chi \bar{\chi}.
\end{align}
Because light mesons are copiously produced at high-energy and very small angles with respect to the beam axis, this primary component provides the irreducible baseline for any far-forward mCP search at the LHC.

For the pseudoscalar channel, we parametrize the branching ratio relative to the $\gamma e^+ e^-$ mode as
\begin{align}
    \frac{\text{BR}(M\to\gamma \chi \bar{\chi})}{\text{BR}(M\to\gamma e^+ e^-)}=\epsilon^2 \frac{I(m_\chi)}{I(m_e)}
\end{align}
where
\begin{align}
    I(m)=\int_{4m^2}^{m_M^2} \dd q^2 \frac{1}{q^2}\left(1-\frac{q^2}{m_M^2}\right)^3\left(1+\frac{2m^2}{q^2}\right)\sqrt{1-\frac{4m^2}{q^2}}
\end{align}
characterizes the 3-body phase space integral.

For the vector-meson channel, the corresponding ratio to the dilepton mode is
\begin{align}
\frac{\mathrm{BR}(M\to\chi\bar{\chi})}{\mathrm{BR}(M\to e^+e^-)}
=
\epsilon^2
\frac{(m_M^2+2m_\chi^2)\,(m_M^2-4m_\chi^2)^{1/2}}
     {(m_M^2+2m_e^2)\,(m_M^2-4m_e^2)^{1/2}}.
\label{eq:vector_mcp}
\end{align}
These expressions determine the probability for a given primary meson to produce an mCP pair once the parent-meson spectrum is specified.

To estimate the IP-originating mCP flux, we simulate forward hadron production in $pp$ collisions using three aforementioned generators: EPOS-LHC, Pythia8 and SIBYLL.\footnote{The QGSJET generator does not support the study of $\rho$ meson production; therefore, we omit it from the subsequent mCP analysis.} For each generated meson, we weight its decay into mCPs by the corresponding branching ratio, sample the daughter kinematics, and boost the final-state particles to the lab frame. 
The produced $\chi$ particles are then propagated from the IP to the benchmark detector introduced in \cref{sec:detection}. Events are retained when at least one mCP trajectory intersects the active detector volume. The corresponding detection probability is evaluated using the ionization formalism described in \cref{sec:detection}, allowing us to determine the expected primary signal yield as a function of $m_\chi$ and $\epsilon$. This analysis utilizes a customized \texttt{FORESEE} package~\cite{Kling:2021fwx} that implements mCP ionization signal detection~\cite{MammenAbraham:2024gun}. 

The characteristic energy distribution of these primary mCPs reaching the detector volume is illustrated in \cref{fig:mCP_spectrum} for a benchmark mass of $m_\chi =50~\textrm{MeV}$. As shown, the spectrum exhibits a prominent peak at approximately $1~\textrm{TeV}$, which reflects the highly boosted nature of parent mesons produced in the far-forward region of the LHC. The distribution also features a significant tail extending toward lower energies. Within the mass range of interest for the analysis below, this primary signal originates predominantly from the decays of $\pi^0$ and $\eta$ mesons.

This IP contribution will be used throughout the rest of the paper as the primary reference signal. In particular, when discussing mCP production from hadronic and electromagnetic showers in the TAXN, we will compare their yields against this baseline in order to quantify the importance of the new secondary source proposed in this work.

\subsection{mCPs from hadronic showers}
\label{sec:mcp_from_hadronic_showers}

We now turn to the new contribution emphasized in this work: mCP production from hadronic showers initiated by forward neutral hadrons in the TAXN absorber. 

The resulting energy distribution of mCPs produced via the decays of secondary mesons in hadronic showers is presented in \cref{fig:mCP_spectrum} for the benchmark mass $m_\chi=50~\textrm{MeV}$. Compared with the primary signal discussed in the previous section, the secondary mCP spectrum is notably shifted toward lower energies. This shift is a characteristic feature of shower development in the TAXN absorber, where the energy of the initiating neutral hadrons is partitioned among multiple generations of secondary particles, resulting in a softer meson spectrum.

However, despite this shift in the energy scale, the high multiplicity of secondary mesons produced in these hadronic cascades compensates for the lower average energy per particle. As a result, the total integrated flux of mCPs from secondary production can be comparable in magnitude to the primary baseline originating directly from the IP. This makes the hadronic shower component a significant contribution to the total mCP signal expected at downstream detectors.

\subsection{mCPs from electromagnetic showers}
\label{sec:EMshower}
\begin{figure}
\centering
\includegraphics[width=\columnwidth]{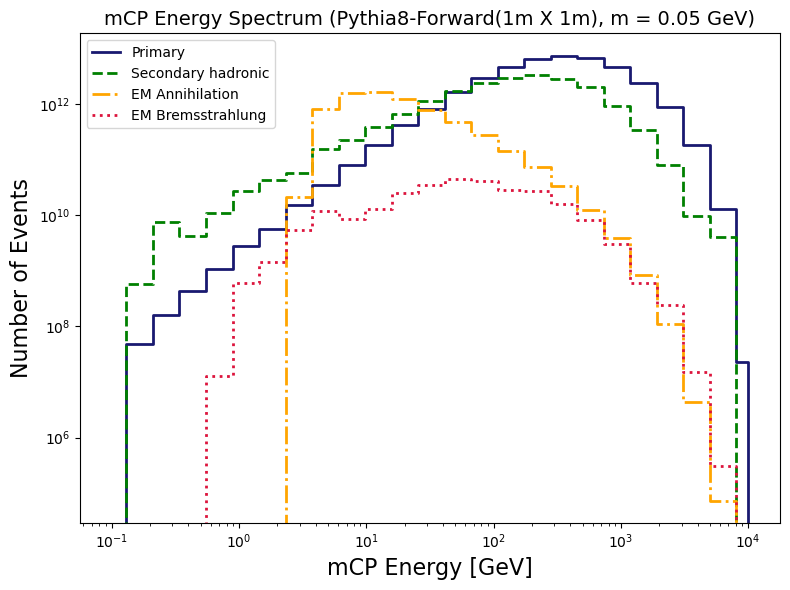}
\caption{Energy spectrum of detected events for mCP mass of 50 MeV for different production channels for Pythia8-Forward generator.}
\label{fig:mCP_spectrum}
\end{figure}
We also include mCP production from the electromagnetic shower initiated in the TAXN absorber. 
While the low-mass mCP flux from the ATLAS IP is dominated by meson decays, the EM cascade provides an additional and non-negligible source of signal events. This is because a large flux of high-energy photons is also produced in the far-forward direction, as discussed in \cref{sec:secondaries}, and initiates an intense EM shower in the absorber.
The two leading EM-shower production channels considered in this work are annihilation and bremsstrahlung. 

{\flushleft \textbf{Annihilation}:}
The large positron multiplicity in the EM shower opens the channel
\begin{align}
e^+ e^- \to \chi \bar{\chi},
\end{align}
through annihilation on bound electrons in the absorber material.  
The corresponding cross section is
\begin{align}
\sigma_{\text{anni.}} &= \frac{4\pi \alpha^2 \varepsilon^2}{3} 
\cdot \frac{(s + 2m_e^2)}{s^3 \sqrt{s - 4m_e^2}} 
\cdot (s + 2m_\chi^2)\sqrt{s - 4m_\chi^2},
\end{align}
where $s = 2m_e E + m_e^2$ denotes the center-of-mass energy squared. 
The number of mCPs from the annihilation production is
\begin{align}
    N = 2\times n_e\, d\, \sigma_{\text{anni.}} \,,
\end{align}
where $n_e$ is the electron number density of the material, and $d$ is the distance traversed by a given positron in the absorber.
For each positron extracted from the \texttt{Geant4} simulation, we evaluate the corresponding production rate and then propagate the produced mCPs to the detector volume. The orange line in \cref{fig:mCP_spectrum} shows the mCP spectrum from the annihilation channel for a benchmark point with $m_\chi = 0.05~\mathrm{GeV}$. It peaks near the kinematic threshold because, in this regime, the $s$-channel production cross section is relatively large, and the $\chi\bar{\chi}$ pair is produced nearly at rest in the center-of-mass frame, making it more easily boosted into the forward direction in the lab frame.

{\flushleft \textbf{Bremsstrahlung}:}
Electrons and positrons in the EM shower can also produce an mCP pair through an off-shell photon emitted in scattering on the absorber material,
\begin{align}
e^\pm N \to e^\pm N \chi \bar{\chi},
\end{align}
where $N$ refers to copper nucleus.
In our analysis, this contribution is computed and simulated with \texttt{MadGraph}, using the information of $e^\pm$ obtained from \texttt{Geant4} as inputs. The contribution from bremsstrahlung production is generally smaller than the 2-to-2 annihilation process (see \cref{fig:mCP_spectrum}).

Although the total flux of both channels remains below the primary IP contribution, their combined contribution can be comparable to that from the neutron-induced hadronic shower. We therefore include the EM-shower contribution as an essential part of the secondary production channels.

\subsection{Millicharged particles and forward detectors}
\label{sec:detection}

As mCPs are DM candidates, they are stable and can be searched for via either ionization or scattering signatures. Since mCPs interact electromagnetically with a suppressed coupling, ionization signatures are generally more sensitive than scattering signals. Ionization arises from the accumulation of many soft Coulomb interactions along the particle's trajectory, leading to a continuous energy deposition that remains detectable even for very small charges, whereas scattering processes correspond to rare hard interactions and are therefore more suppressed. The energy deposition rate, $\left\langle -\frac{dE}{dx} \right\rangle$, is given by the Bethe--Bloch formula~\cite{ParticleDataGroup:2024cfk}. The average number of photoelectrons collected per layer of the detector is given by
\begin{equation}
    N_{\mathrm{pe}} =  \rho \ \left\langle -\frac{dE}{dx} \right\rangle \times D_{\mathrm{eff}} \times P_{\mathrm{yield}} \times L_{\mathrm{det}},
\end{equation}
where $\rho$ is the mass density of the detector material, $D_{\mathrm{eff}}$ is the detector efficiency, $P_{\mathrm{yield}}$ is the number of photons per MeV of energy deposited in the material, and $L_{\mathrm{det}}$ is the length of the detector layer. The probability of detection is given by

\begin{equation}
    \mathcal{P}_{\mathrm{det}} = (1 - e^{-N_\mathrm{pe}})^{n_\mathrm{layers}},
\end{equation}
where $n_\mathrm{layers}$ is the number of layers of the detector. The FORMOSA detector considers a BC-408 plastic scintillator detector with an effective volume of $1 \ \mathrm{m}\times1 \ \mathrm{m}\times 4\ \mathrm{m}$ consisting of four layers with each layer being $1\ \mathrm{m}$ in length. The photon yield is $P_{\mathrm{yield}} = 11.14 \times 10^{3} \ \mathrm{MeV^{-1}}$ 
and the density of material is $1.023 \ \mathrm{g/cm}^3$. We take the detector efficiency to be $D_{\mathrm{eff}}=10\%$. Alternative detector materials with higher photon yields, such as $\mathrm{CeBr_3}$, have recently been proposed, which could lead to improved sensitivity in the low-mass and low-charge regime of the mCP parameter space~\cite{Citron:2025kcy}. We have verified that the relative contribution from secondary-induced showers remains largely unchanged with different materials. Therefore, we proceed with the plastic scintillator as our baseline choice when discussing the comparison between the primary and secondary mCP production efficiency.

Within the energy range relevant for the forward mCP flux in this work, the detection probability depends only mildly on the mCP energy, as the mCPs reaching the detector are typically relativistic. Consequently, the low-energy tail of the mCP spectrum can make a sizable contribution to the total event yield due to the high particle multiplicity. As illustrated in \cref{fig:mCP_spectrum}, this tail is particularly pronounced for mCPs originating from EM showers, while it is more suppressed for the characteristically harder mCPs produced in hadronic showers.

These energy differences are important when considering how downstream LHC magnets affect the forward mCP flux. The design of the superconducting coils and the surrounding iron yoke is optimized to confine the strong magnetic field ($B \approx 8.33~\text{T}$) within the two separate beam channels in which protons travel in opposite directions~\cite{Bruning:2004ej}. As a result, mCPs whose trajectories lie outside these apertures experience a weaker field. However, mCPs that pass directly through the high-field regions inside the pipes will experience a stronger transverse Lorentz force, leading to a deflection that is inversely proportional to the particle's momentum. While this effect is suppressed for the characteristically harder mCPs from hadronic showers, the softer mCPs originating from EM showers are more likely to be deflected, especially for larger values of $\epsilon$.

Interestingly, these fields do not only act to suppress the flux. As studied in the context of the proposed sweeper magnet for the FPF~\cite{Feng:2022inv,FPF:2025bor}, forward LHC magnets can deflect particles both away from and toward the detector's axis. Similarly, downstream LHC magnets could potentially help focus a portion of the secondary mCP flux toward the FORMOSA detector. A precise assessment of these effects would require a detailed map of the magnetic field and a full particle-tracking simulation. While such modeling is beyond the scope of this study, it remains an important requirement for future high-precision flux predictions, particularly for low-momentum particles and higher charges. In the following, we shall assume that these magnetic effects are negligible, which is expected to be a valid approximation particularly for the dominant secondary mCP contribution from hadronic showers and for low $\epsilon$ close to the projected FORMOSA sensitivity bounds.

\section{Results}
\label{sec:results}
\begin{figure}
\centering
\includegraphics[width=\columnwidth]{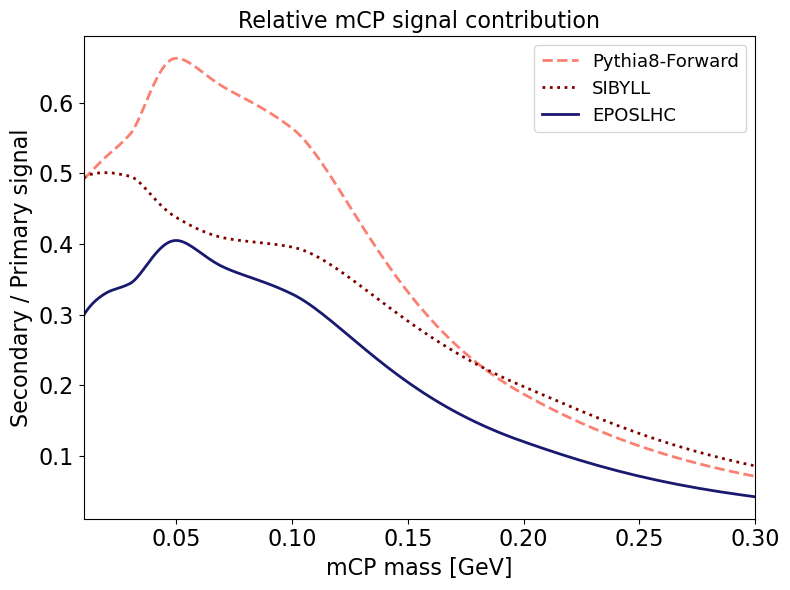}
\caption{Comparison of secondary signal importance for three different generators for detector configuration of $1\,\mathrm{m} \times 1\,\mathrm{m} \times 4\,\mathrm{m}$ 
}
\label{fig:Gen_compare}
\end{figure}
In \cref{fig:Gen_compare}, we present the ratio of the secondary mCP signal, summing hadronic and electromagnetic shower components, to the signal from primary IP-produced mCPs as a function of the mCP mass, $m_\chi$. The results represent the ratio of detected signal events within the $1~\textrm{m}\times 1~\textrm{m}\times 4~\textrm{m}$ detector geometry. As shown, the secondary contribution is significant: for the EPOS-LHC generator, the secondary-to-primary ratio reaches approximately $40\%$, while for the forward-tuned Pythia 8, it exceeds $60\%$.

For these two generators, the ratio grows steadily with $m_\chi$ up to approximately $0.05~\textrm{GeV}$. In this regime, while pion decays remain the dominant source of mCPs, the contribution from $\eta$ mesons begins to emerge. The relative efficiency of $\eta$ production appears lower in secondary showers compared to primary $pp$ interactions, likely due to the lower average energy available per interaction within the cascade; this leads to a diminishing relative secondary contribution as the mCP mass increases. In contrast, the SIBYLL generator lacks a distinct peaking structure in this region, which we attribute to its higher predicted rate of primary $\eta$ production compared to the other two generators, thereby increasing the primary baseline. For masses above $0.2-0.3~\textrm{GeV}$, we account for the contribution from $\rho$ mesons; however, the secondary ratio generally drops below $10\%$ in this mass range. These results indicate that the impact of secondary shower production is most substantial for $m_\chi\lesssim 0.2~\textrm{GeV}$, justifying our focus on light mesons and the omission of heavier states for the secondary flux calculations.

\begin{figure}
\centering
\includegraphics[width=\columnwidth]{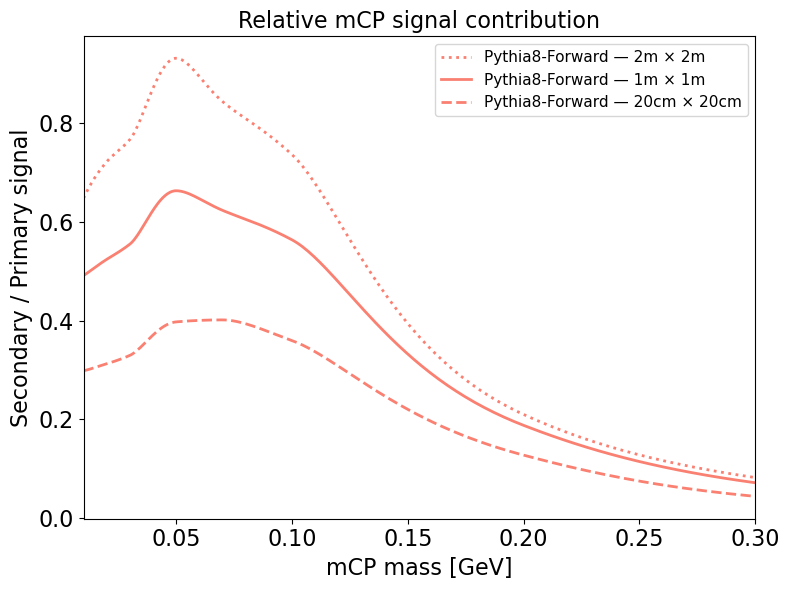}
\caption{Comparison of secondary signal importance for three different detector geometries for Pythia8-Forward generator.}
\label{fig:Geom_compare}
\end{figure}

To quantify how the impact of secondary showers depends on the angular acceptance, we show in \cref{fig:Geom_compare} the ratio of secondary-to-primary mCP event rates for three different transverse detector geometries: $20~\textrm{cm}\times 20~\textrm{cm}$, $1~\textrm{m}\times 1~\textrm{m}$, and $2~\textrm{m}\times 2~\textrm{m}$. The plot was obtained using Pythia 8 forward generator. As expected, the relative importance of the secondary component increases with larger detector sizes. This is because secondary production within the TAXN occurs over a broader range of angles compared to the highly collimated primary flux originating from the IP. While the active scintillator volume of the proposed FORMOSA detector is planned to be $1~\textrm{m}\times 1~\textrm{m}$, the detector modules are expected to be distributed across a wider $1.9~\textrm{m}\times 1.9~\textrm{m}$ array~\cite{Citron:2025kcy}. In this larger transverse regime, the secondary contribution becomes even more dominant, potentially enhancing the signal by up to $90\%$ relative to the primary baseline and stressing the importance of accounting for secondary shower production in realistic detector geometries.

\begin{figure}
\centering
\includegraphics[width=\columnwidth]{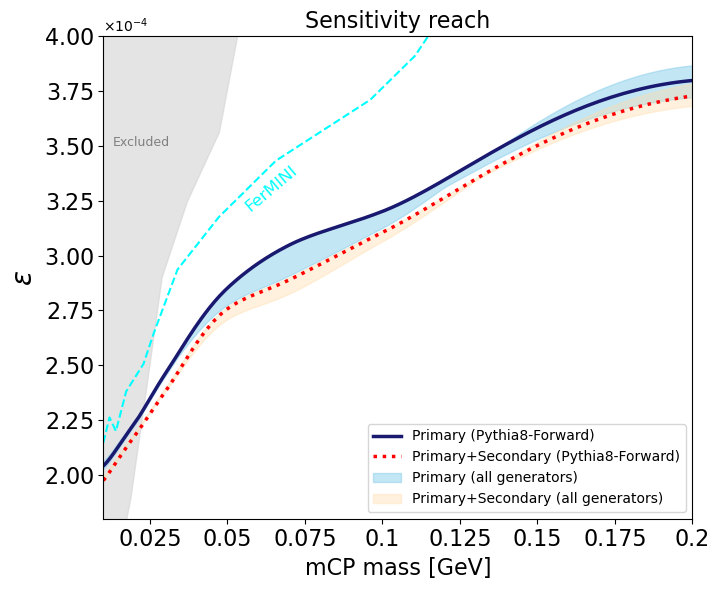}
\caption{{The projected sensitivity of FORMOSA to mCPs produced from primary hadrons is shown by the solid blue line, with the surrounding blue band indicating the uncertainty arising from different production generators. The red dotted line represents the sensitivity when both primary and secondary shower contributions are included, while the corresponding orange band reflects the associated generator uncertainties. The gray shaded region denotes the parameter space already excluded by existing constraints, and the dotted cyan line illustrates the projected sensitivity of FerMINI \cite{Kelly:2018brz}. }}
\label{fig:Sensitivity_reach}
\end{figure}

\cref{fig:Sensitivity_reach} shows the projected exclusion bound in the kinetic mixing parameter, $\epsilon$, as a function of the mCP mass. For comparison, the dashed cyan curve indicates the projected reach of the FerMINI experiment \cite{Kelly:2018brz}, and the gray shaded region denotes the parameter space already excluded by existing constraints. The solid blue curve represents the projected FORMOSA bound obtained from primary production only, simulated using Pythia 8 in the forward configuration, while the red dotted curve illustrates the total reach when both primary and secondary production channels are included. To account for modeling choices, we provide colored shaded bands indicating the theoretical uncertainty on the projected exclusion bounds derived from the different forward MC generators discussed in \cref{sec:secondaries}. Crucially, the shift in sensitivity resulting from secondary production exceeds the width of the uncertainty band for the primary-only prediction in the low-mass regime. This demonstrates that the impact of secondary showering is a dominant effect that is not obscured by current theoretical modeling uncertainties, and is therefore essential for a realistic assessment of the detector's potential.

\section{Conclusion}
\label{sec:conclusion}

In this work, we identify and characterize a new source of millicharged particles at the LHC: secondary production in showers initiated by far-forward neutral particles in the TAXN absorber. While previous studies of forward mCP searches have primarily focused on direct production at the ATLAS interaction point, we have shown that the LHC can also effectively operate as a beam-dump setup, in which energetic forward neutrons and photons generate substantial hadronic and electromagnetic cascades downstream. These secondary showers produce additional light mesons and energetic $e^\pm$, which in turn yield an extra flux of forward-going mCPs.

Using Monte Carlo simulations of forward particle production together with \texttt{Geant4} showering in the neutral absorber TAXN, we have quantified both the primary and secondary contributions to the mCP signal at a far-forward detector. We find that neutron-induced hadronic showers provide a sizable additional source of mCPs, especially in the low-mass region, where secondary $\pi^0$, $\eta$, and $\rho^0$ decays remain efficient. We also find that the contribution from electromagnetic showers can be comparable to that from hadronic showers. Taken together, these secondary production channels can enhance the expected signal yield by about $50-60\%$ relative to the primary production alone for $m_\chi \lesssim 0.1~\mathrm{GeV}$, assuming a benchmark FORMOSA detector geometry with transverse size $1~\mathrm{m}\times 1~\mathrm{m}$. Our results show that incorporating secondary production is essential for a proper assessment of exclusion bounds and related theoretical uncertainties in future far-forward searches for mCPs at the LHC.

Ultimately, secondary production in downstream absorbers constitutes an additional production mode that should not be overlooked in searches for new physics at the LHC. The framework developed here for millicharged particles is readily generalizable to a wide range of light, weakly coupled BSM scenarios. To facilitate such future investigations and to aid in the refinement of forward signal simulations, we provide the data for our forward meson and electromagnetic shower spectra in a public repository. Integrating these secondary sources into the standard simulation chain will contribute to maximizing the scientific output of the HL-LHC far-forward physics program.

\begin{acknowledgments}
We thank Felix Kling for helpful discussions and comments on the manuscript. We thank Matthew Citron, Max Fieg, and Juan Tafoya Vargas for useful discussions. J.A. and S.T. are supported by the National Science Centre, Poland (research grant No. 2021/42/E/ST2/00031). S.T. and A.Z. are supported by Teaming for Excellence grant Astrocent Plus (GA: 101137080) funded by the European Union, with complementary national funding from the MNiSW (MNiSW/2025/DIR/811). The work of A.Z. was supported by the International Research Agenda Programmes of the Foundation for Polish Science: AstroCeNT (MAB/2018/7), funded from the European Regional Development Fund, and Astrocent (FENG.02.01-IP.05-A015/25), co-financed by the European Union under the European Funds for Smart Economy 2021-2027 (FENG). A.Z. was also supported by the National Science Centre, Poland (research grants No. 2021/42/E/ST2/00331 and 2025/09/X/ST2/00794). P.L. and Z.L. are supported by the Department of Energy under Grant No.~DE-SC0011842 at the University of Minnesota. P.L. is partly supported by a Doctoral Dissertation Fellowship at the University of Minnesota. Z.L. is supported in part by a Sloan Research Fellowship from the Alfred P. Sloan Foundation at the University of Minnesota.  
\end{acknowledgments}

\bibliography{ref}
\end{document}